\documentclass[aps,prd,nofootinbib,superscriptaddress,11pt]{revtex4}
\usepackage{txfonts}
\usepackage{graphicx}
\usepackage{dcolumn}
\usepackage{bm}
\usepackage{amssymb}
\usepackage{latexsym}

\newcommand{\be}{\begin{equation}}
\newcommand{\ee}{\end{equation}}
\newcommand{\bq}{\begin{eqnarray}}
\newcommand{\eq}{\end{eqnarray}}

\begin{document}

\title{Fitting the Constitution SNIa Data with Redshift Binned Parameterization Method}

\author{Qing-Guo Huang}
\email{huangqg@kias.re.kr} \affiliation{School of Physics, Korea
Institute of Advanced Study, Seoul 130722, Korea} \affiliation{Kavli
Institute for Theoretical Physics China, Chinese Academy of
Sciences, Beijing 100190, China}
\author{Miao Li}
\email{mli@itp.ac.cn} \affiliation{Kavli Institute for Theoretical
Physics China, Chinese Academy
of Sciences, Beijing 100190, China}
\affiliation{Key Laboratory of Frontiers in Theoretical Physics,
Institute of Theoretical Physics, Chinese Academy
of Sciences, Beijing 100190, China}

\author{Xiao-Dong Li}
\email{renzhe@mail.ustc.edu.cn} \affiliation{Interdisciplinary
Center for Theoretical Study, University of Science and Technology
of China, Hefei 230026, China} \affiliation{Key Laboratory of Frontiers in Theoretical Physics,
Institute of Theoretical
Physics, Chinese Academy of Sciences, Beijing 100190, China}

\author{Shuang Wang}
\email{swang@mail.ustc.edu.cn} \affiliation{Department of Modern
Physics, University of Science and Technology of China, Hefei
230026, China} \affiliation{Key Laboratory of Frontiers in Theoretical Physics,
Institute of Theoretical Physics,
Chinese Academy of Sciences, Beijing 100190, China}

\begin{abstract}

In this work, we explore the cosmological consequences of the
recently released Constitution sample of 397 Type Ia supernovae
(SNIa). By revisiting the Chevallier-Polarski-Linder (CPL)
parameterization, we find that, for fitting the Constitution set
alone, the behavior of dark energy (DE) significantly deviate from
the cosmological constant $\Lambda$, where the equation of state
(EOS) $w$ and the energy density $\rho_{\Lambda}$ of DE will rapidly
decrease along with the increase of redshift $z$. Inspired by this
clue, we separate the redshifts into different bins, and discuss the
models of a constant $w$ or a constant $\rho_{\Lambda}$ in each bin,
respectively. It is found that for fitting the Constitution set
alone, $w$ and $\rho_{\Lambda}$ will also rapidly decrease along
with the increase of $z$, which is consistent with the result of CPL
model. Moreover, a step function model in which $\rho_{\Lambda}$
rapidly decreases at redshift $z\sim0.331$ presents a significant
improvement ($\Delta \chi^{2}=-4.361$) over the CPL
parameterization, and performs better than other DE models. We also
plot the error bars of DE density of this model, and find that this
model deviates from the cosmological constant $\Lambda$ at 68.3\%
confidence level (CL); this may arise from some biasing systematic
errors in the handling of SNIa data, or more interestingly from the
nature of DE itself. In addition, for models with same number of
redshift bins, a piecewise constant $\rho_{\Lambda}$ model always
performs better than a piecewise constant $w$ model; this shows the
advantage of using $\rho_{\Lambda}$, instead of $w$, to probe the
variation of DE.

\end{abstract}

\maketitle

\section{Introduction}\label{sec:intro}

Although it has been a decade since the discovery of the cosmic acceleration \cite{Riess,Perlmutter},
the nature of dark energy (DE) still remains a mystery.
The most obvious theoretical candidate of DE is the cosmological constant $\Lambda$,
which can fit observations well, but is plagued with the fine-tuning problem and the coincidence problem \cite{Weinberg}.
Numerous other dynamical DE models have also been proposed in the literature,
such as quintessence \cite{quint}, phantom \cite{phantom}, $k$-essence \cite{k}, tachyon \cite{tachyonic},
holographic \cite{holographic}, agegraphic \cite{agegraphic},
hessence \cite{hessence}, Chaplygin gas \cite{Chaplygin}, Yang-Mills condensate \cite{YMC}, ect.

A most powerful probe of DE is Type Ia supernovae (SNIa), which can
be used as cosmological standard candles to measure directly the
expansion history of the universe. A large sample of nearby SnIa
with $z<0.08$ has recently been published \cite{Hicken1}. Adding
this to the Union sample \cite{Kowalski} leads to the so-called
Constitution set \cite{Hicken2} which is currently the largest SnIa
sample to date. By analyzing this Constitution set with
Chevallier-Polarski-Linder (CPL) ansatz \cite{CPL}, Shafieloo et al.
\cite{Shafieloo} argue that cosmic acceleration may have already
peaked and that we are currently witnessing its slowing down. By
separating the redshifts into different bins and assuming a constant
equation of state (EOS) $w$ in each bin, Qi et al. \cite{Qi} also
point out that this Constitution set shows a deviation at 68.3\%
confidence level (CL) from the $\Lambda$CDM model. Besides, by
studying the constraints of DE on the $w$ - $w'$ plane, Chen et al.
\cite{Chen} claim that $\Lambda$CDM model is disfavored by the
current observational data at 68.3\% confidence.

Although constraining EOS $w$ of DE is a popular and widely-used
method to investigate dark energy, Wang and Freese \cite{ywang1}
pointed out that DE density $\rho_{\Lambda}$ can be constrained more
tightly than EOS $w$ given the same observational data. Therefore,
it would also be very interesting to consider the cases of piecewise
constant $\rho_{\Lambda}$. In this work we separate the redshifts
into different bins, and discuss the models of a constant $w$ or a
constant $\rho_{\Lambda}$ in each bin, respectively. Binned fits of
EOS have been applied  before in \cite{Huterer} \cite{Sullivan}
\cite{Qi2}, and the Union paper itself \cite{Kowalski}.  Similar
analyses have been performed for density binning \cite{ywang1}. Our
paper differs from previous ones in two aspects: First, in previous
papers the redshift bins are determined by setting the discontinuity
points of redshift by hand, while in our paper we treat the
discontinuity points of redshift as free parameters, and find that
much smaller $\chi_{min}^{2}$ can be obtained. Second, not only the
piecewise constant $w$ models but also the piecewise constant
$\rho_{\Lambda}$ models are considered in this work. Moreover, by
comparing these two classes of model, it is found that for same
number of redshift bins, a piecewise constant $\rho_{\Lambda}$ model
always performs better than a piecewise constant $w$ model.

This work is organized as follows. In Section 2, we briefly review
the models considered here and the method of data analysis. In
Section 3, we introduce the observational data and describe how they
are included in our analysis. Section 4 is divided into three parts:
First of all, we revisit the CPL model and find that for fitting the
Constitution set alone the behavior of DE significantly deviate from
the cosmological constant $\Lambda$, where $w$ and $\rho_{\Lambda}$
will rapidly decrease along with the increase of redshift $z$. This
result implies that the behavior of dark energy might be very
different in different slices of redshifts, and inspires us to
separate redshifts into several bins and to investigate the
piecewise constant $w$ model and the piecewise constant
$\rho_{\Lambda}$ model. Then, we discuss the models with two bins.
It is found that for fitting the Constitution set alone,
$\rho_{\Lambda}$ will also rapidly decrease along with the increase
of $z$, which is consistent with the result of CPL model. Moreover,
a step function model in which $\rho_{\Lambda}$ rapidly decreases at
redshift $z\sim0.331$ presents a significant improvement ($\Delta
\chi^{2}=-4.361$) over the CPL parameterization, and performs better
than other DE models. We also plot the error bars of
$\rho_{\Lambda}$ of this model, and find that this model deviates
from the cosmological constant $\Lambda$ at 68.3\% confidence level.
Next, we briefly discuss the models with three bins, and find that
the cases of three bins are very similar with that of two bins. At
last, we give a short summary in Section 5. In this work, we assume
today's scale factor $a_{0}=1$, so the redshift $z$ satisfies
$z=a^{-1}-1$; the subscript ``0'' always indicates the present value
of the corresponding quantity, and the unit with $c=\hbar=1$ is
used.

\section{Model And Methodology}

Standard candles impose constraints on cosmological parameters essentially
through a comparison of the luminosity distance from observation with that from theoretical models.
In a spatially flat Friedmann-Robertson-Walker (FRW) universe
(the assumption of flatness is motivated by the inflation scenario),
the luminosity distance $d_L$ is given by
\begin{equation}
\label{eq:dl}
d_L(z)=\frac{1+z}{H_{0}}\int_0^z\frac{dz'}{E(z')},
\end{equation}
with
\begin{equation}
\label{eq:Ez}
E(z)\equiv H(z)/H_{0} =\left[\Omega_{m0}(1+z)^3+(1-\Omega_{m0})f(z)\right]^{1/2},
\end{equation}
where $H(z)$ is the Hubble parameter, $H_{0}$ is the Hubble constant,
$\Omega_{m0}$ is the present fractional matter density,
and $f(z)\equiv \rho_{\Lambda}(z)/\rho_{\Lambda 0}$ is a key function,
because DE parameterization schemes enter through $f(z)$.

We will consider the familiar CPL ansatz \cite{CPL}, in which the EoS of DE is parameterized as
\begin{equation}
\label{eq:cpl1}
w=w_0+w_1\frac{z}{1+z}\,,
\end{equation}
where $w_0$ and $w_1$ are constants. As is well known, the corresponding $f(z)$ is given by
\begin{equation}
\label{eq:cpl2}
f(z)=(1+z)^{3(1+w_0+w_1)}\exp\left(-\frac{3w_1 z}{1+z}\right).
\end{equation}
For the case where EOS $w$ is piecewise constant in redshift,
$f(z)$ can be written as \citep{Sullivan,Qi2}
\begin{equation}
\label{eq:fzwbinned}
f(z_{n-1}<z \le z_n)=(1+z)^{3(1+w_n)}\prod_{i=0}^{n-1}(1+z_i)^{3(w_i-w_{i+1})},
\end{equation}
where $w_i$ is the EOS parameter in the $i^{th}$ redshift bin defined by an upper boundary at $z_i$.
This class of models has been extensively studied in the literature.
As mentioned above, it is interesting to consider the piecewise constant $\rho_{\Lambda}$ models.
For this case, $f(z)$ can be written as
\begin{equation}
\label{eq:fzrbinned}
f(z)=
\left\{
\begin{array}{ll}
1 & 0\leq z \leq z_{1}
\\
\epsilon_{n} & z_{n-1} \leq z \leq z_{n} ~ (n>1)
\end{array}
\right..
\end{equation}
Here $\epsilon_{n}$ is a a piecewise constant,
and from the relation $E(0)=1$ one can easily obtain $\epsilon_{1}=1$.
So for same number of redshift bins,
the number of free parameters of piecewise constant $\rho_{\Lambda}$ model is one fewer than that of piecewise constant $w$ model.
It should be mentioned that there are different opinions in the literature
about the optimal choice of redshift bins in constraining DE.
In \cite{Sullivan,Qi2}, the authors directly set the discontinuity points of redshift as $z_{1}=0.2$, $z_{2}=0.5$, and $z_{3}=1.8$.
In \cite{ywang2}, Wang argue that one should choose a constant $\Delta z$ for redshift slices.
In this work, we just treat the discontinuity points of redshift as model parameters in performing the best-fit analysis.
As seen below, this leads to a much smaller $\chi_{min}^{2}$.

In this work we adopt $\chi^2$ statistic to estimate parameters.
For a physical quantity $\xi$ with experimentally measured value $\xi_o$,
standard deviation $\sigma_{\xi}$, and theoretically predicted value $\xi_t$,
the $\chi^2$ value is given by
\begin{equation}
\label{eq:chi2_xi}
\chi_{\xi}^2=\frac{\left(\xi_t-\xi_o\right)^2}{\sigma_{\xi}^2}.
\end{equation}
The total $\chi^2$ is the sum of all $\chi_{\xi}^2$s,
i.e.
\begin{equation}
\label{eq:chi2}
\chi^2=\sum_{\xi}\chi_{\xi}^2.
\end{equation}
Assuming the measurement errors be Gaussian,
the likelihood function is given by
\begin{equation}
\label{eq:likelihood}
{\cal{L } } \propto e^{-\chi^2/2}.
\end{equation}
For comparing different models, a statistical variable must be chosen.
The $\chi _{min}^{2}$ is the simplest one,
but it has difficulty to compare different models with different number of parameters.
In this work, we will use $\chi _{min}^{2}/dof$ as a model selection criterion,
where $dof$ is the degree of freedom defined as
\begin{equation}
\label{eq:dof}
dof\equiv N-k,
\end{equation}
here $N$ is the number of data, and $k$ is the number of free parameters.
Besides, to compare different models with different number of parameters,
people often use the Bayesian information criterion \cite{Liddle} given by \cite{Schwarz}
\begin{equation}
\label{eq:BIC}
BIC=-2 \ln {\cal{L }}_{max}+k \ln N,
\end{equation}
where ${\cal{L }}_{max}$ is the maximum likelihood.
It is clear that a model favored by the observations should give smaller $\chi _{min}^{2}/dof$ and $BIC$.

\section{Observational data}

\subsection{Type Ia supernovae}

For SNIa data, we use the latest 397 Constitution sample,
the distance modulus $\mu_{ obs}(z_i)$, compiled in Table 1 of \cite{Hicken2}.
The theoretical distance modulus is defined as
\begin{equation}
\mu_{th}(z_i)\equiv 5 \log_{10} {D_L(z_i)} +\mu_0,
\end{equation}
where $\mu_0\equiv 42.38-5\log_{10}h$ with $h$ the Hubble constant $H_0$ in units of 100 km/s/Mpc, and
\begin{equation}
D_L(z)=(1+z)\int_0^z {dz'\over E(z';{\bf \theta})}
\end{equation}
is the Hubble-free luminosity distance $H_0d_L$ in a spatially flat FRW universe,
and here ${\bf \theta}$ denotes the model parameters.
The $\chi^2$ for the SNIa data is
\begin{equation}
\chi^2_{SN}({\bf\theta})=\sum\limits_{i=1}^{397}{[\mu_{obs}(z_i)-\mu_{th}(z_i)]^2\over \sigma_i^2},\label{ochisn}
\end{equation}
where $\mu_{obs}(z_i)$ and $\sigma_i$ are the observed value and the corresponding 1$\sigma$ error of distance modulus for each supernova, respectively.
The parameter $\mu_0$ is a nuisance parameter but it is independent of the data and the dataset.
Following \cite{Nesseris}, the minimization with respect to $\mu_0$ can be made trivial by expanding the $\chi^2$ of Eq.(\ref{ochisn}) with respect to $\mu_0$ as
\begin{equation}
\chi^2_{SN}({\bf\theta})=A({\bf\theta})-2\mu_0
B({\bf\theta})+\mu_0^2 C,
\end{equation}
where
\begin{equation}
A({\bf\theta})=\sum\limits_{i=1}^{397}{[\mu_{obs}(z_i)-\mu_{th}(z_i;\mu_0=0,{\bf\theta})]^2\over
\sigma_i^2},
\end{equation}
\begin{equation}
B({\bf\theta})=\sum\limits_{i=1}^{397}{\mu_{obs}(z_i)-\mu_{th}(z_i;\mu_0=0,{\bf\theta})\over
\sigma_i^2},
\end{equation}
\begin{equation}
C=\sum\limits_{i=1}^{397}{1\over \sigma_i^2}.
\end{equation}
Evidently, Eq. (\ref{ochisn}) has a minimum for $\mu_0=B/C$ at
\begin{equation}
\tilde{\chi}^2_{
SN}({\bf\theta})=A({\bf\theta})-{B({\bf\theta})^2\over
C}.\label{tchi2sn}
\end{equation}
Since $\chi^2_{SN, min}=\tilde{\chi}^2_{SN, min}$,
instead minimizing $\chi^2_{SN}$ one can minimize $\tilde{\chi}^2_{SN}$ which is independent of the nuisance parameter $\mu_0$.

\subsection{Baryon Acoustic Oscillations}

For BAO data, we consider the parameter $A$ from the measurement of the BAO peak in the distribution of SDSS luminous red galaxies,
which is defined as \cite{Eisenstein}
\begin{equation}
A\equiv \Omega_{m0}^{1/2} E(z_{b})^{-1/3}\left[{1\over z_{b}}\int_0^{z_{b}}{dz'\over E(z')}\right]^{2/3},
\end{equation}
where $z_{b}=0.35$. The SDSS BAO measurement \cite{Eisenstein} gives $A_{obs}=0.469\,(n_s/0.98)^{-0.35}\pm 0.017$,
where the scalar spectral index is taken to be $n_s=0.960$ as measured by WMAP5 \cite{Komatsu}.
It is widely believed that $A$ is nearly model-independent and can provide robust constraint as complement to SNIa data.
The $\chi^2$ for the BAO data is
\begin{equation}\label{chiLSS}
\chi^2_{BAO}=\frac{(A-A_{obs})^2}{\sigma_A^2},
\end{equation}
where the corresponding $1\sigma$ errors is $\sigma_A=0.017$.

\subsection{cosmic microwave background}

For CMB data, we use the CMB shift parameter $R$, which is given by \cite{Bond,ywang3}
\begin{equation}
R\equiv \Omega_{m0}^{1/2}\int_0^{z_{rec}}{dz'\over E(z')},
\end{equation}
where the redshift of recombination $z_{rec}=1090$ \cite{Komatsu}.
The shift parameter $R$ relates the angular diameter distance to the last scattering surface, the comoving size of the sound horizon at $z_{rec}$
and the angular scale of the first acoustic peak in CMB power spectrum of temperature fluctuations \cite{Bond,ywang3}.
The measured value of $R$ has been updated to be $R_{obs}=1.710\pm 0.019$ from WMAP5 \cite{Komatsu}.
It should be noted that, different from the SNIa and BAO data,
the $R$ parameter can provide the information about the universe at very high redshift.
The $\chi^2$ for the CMB data is
\begin{equation}\label{chiCMB}
\chi^2_{CMB}=\frac{(R-R_{obs})^2}{\sigma_R^2},
\end{equation}
where the corresponding $1\sigma$ errors is $\sigma_R=0.019$.

\section{Results}

\subsection{CPL parameterization}

\begin{figure}
\includegraphics[scale=0.29, angle=0]{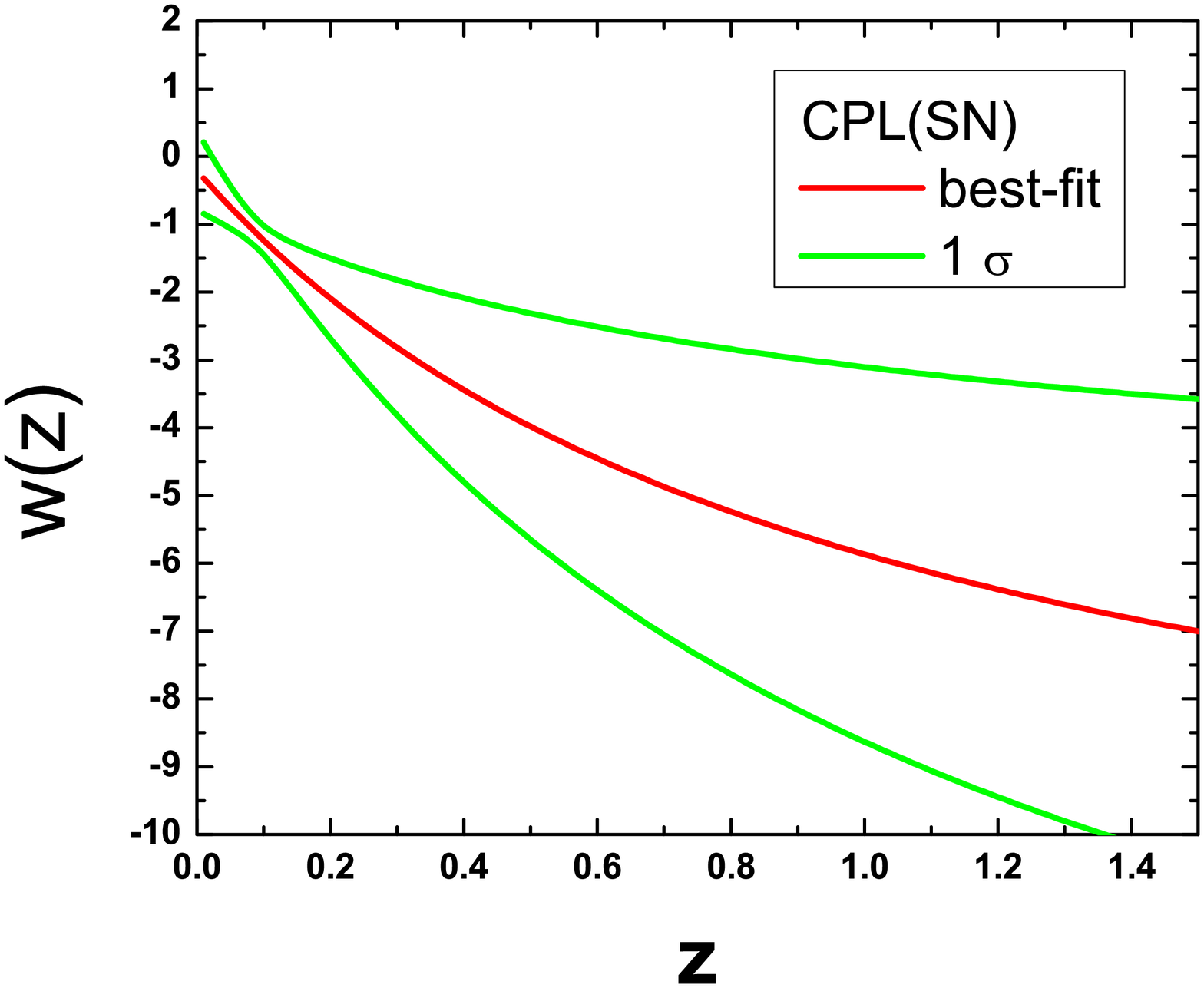}
\includegraphics[scale=0.29, angle=0]{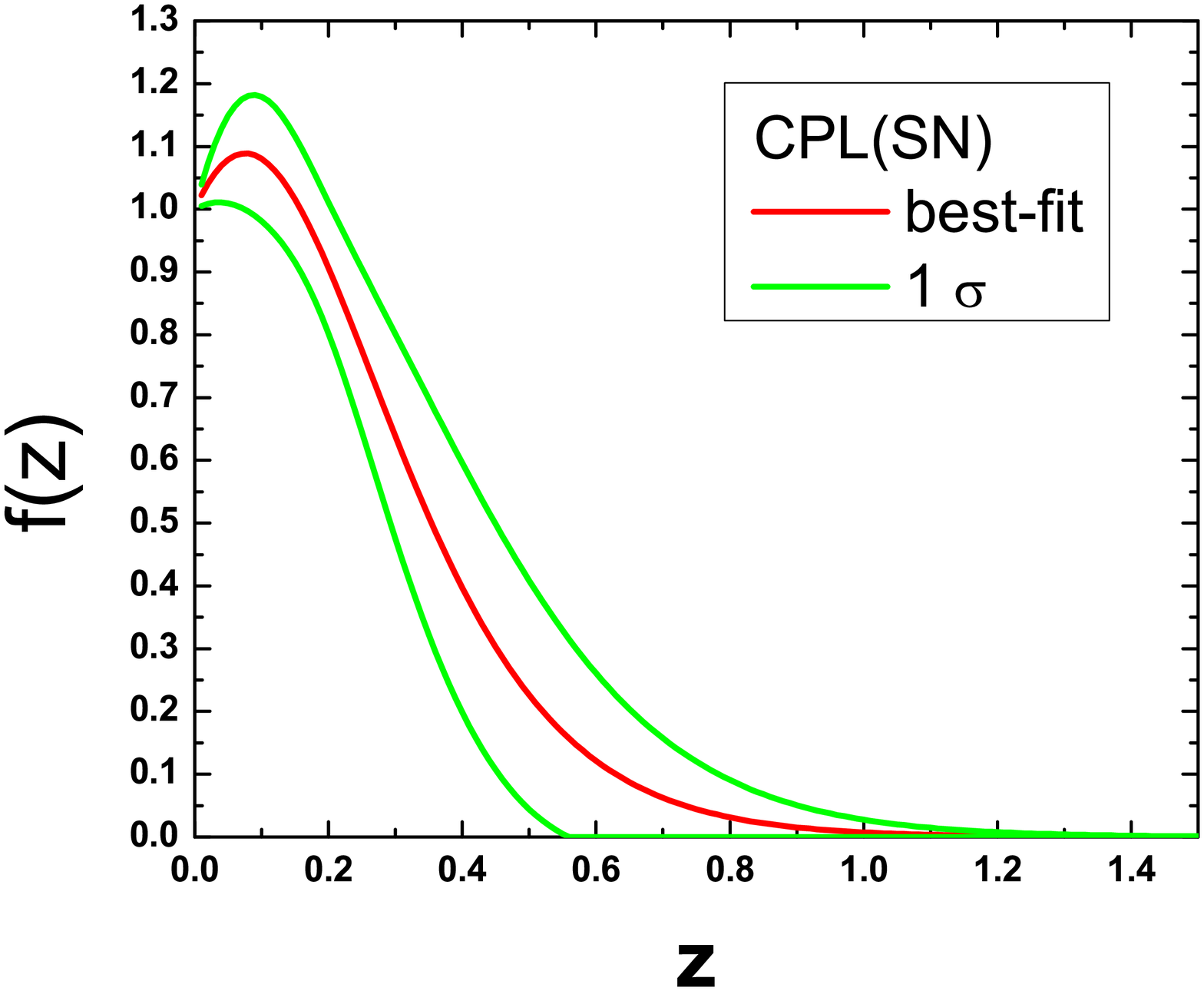}
\includegraphics[scale=0.29, angle=0]{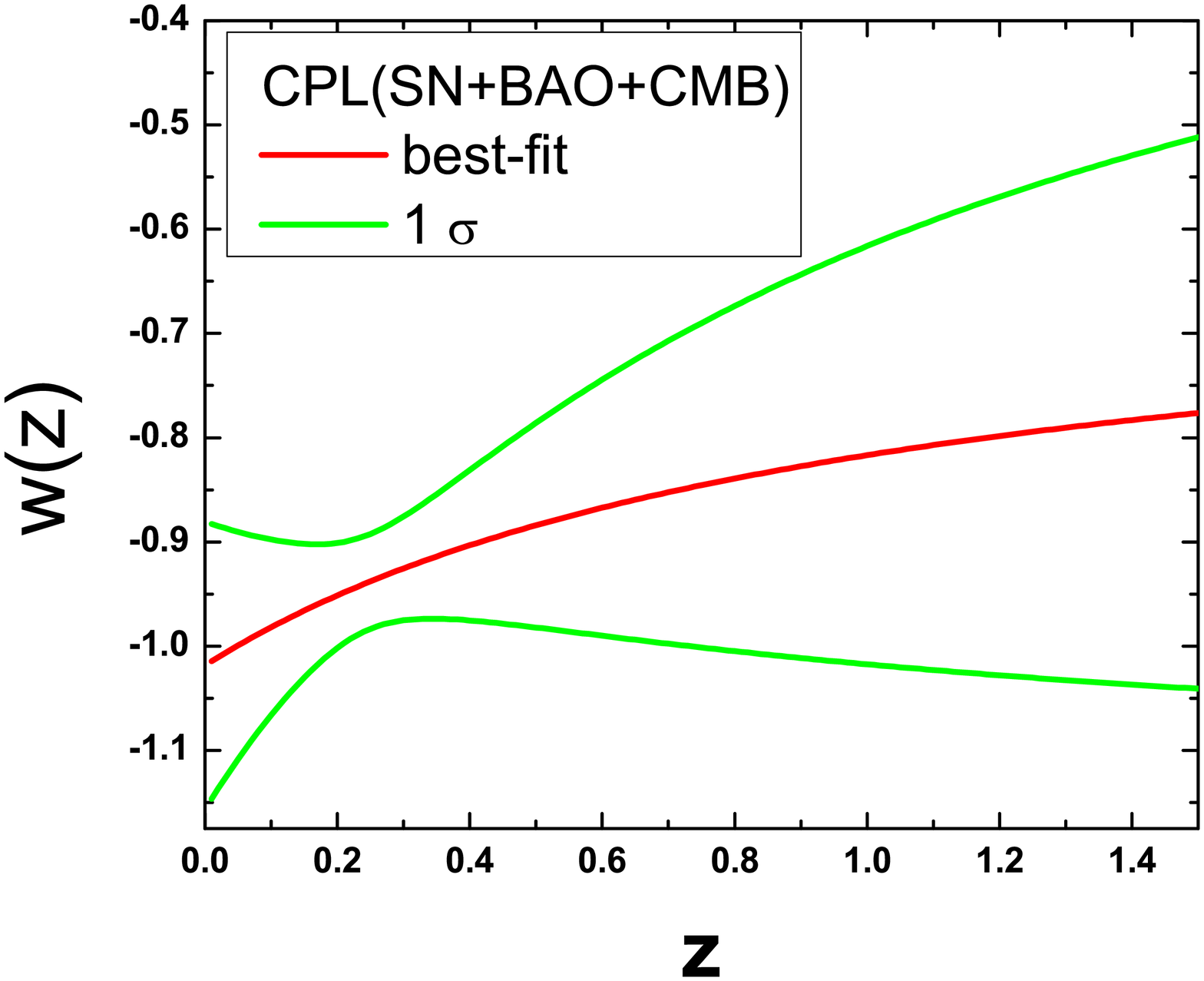}
\includegraphics[scale=0.29, angle=0]{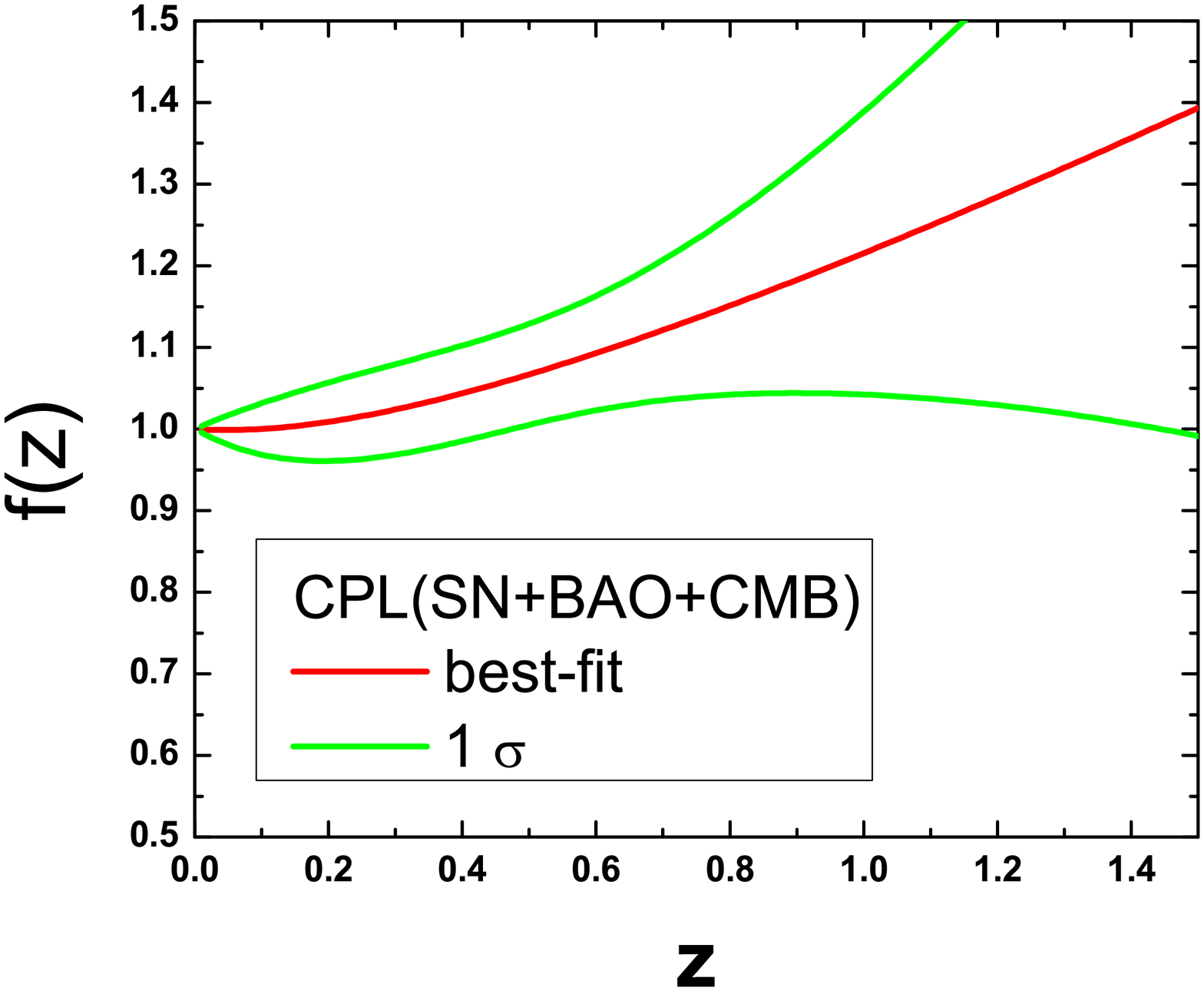}
\caption{\label{fig-CPL} Reconstructed $w(z)$ and $f(z)$ from SnIa
data alone (upper panel) and SnIa+BAO+CMB data(lower panel) using
the CPL ansatz. Red lines stand for the best-fit values and green
lines stand for the $1\sigma$ CL.}

\end{figure}

We will present the results of CPL parameterization in this subsection.
It is found that $\chi_{min}^{2}=461.254$, $465.440$, and $466.100$ for SNIa, SNIa+BAO, and SNIa+BAO+CMB data, respectively.
Moreover, we reconstructe $w(z)$ and $f(z)$ by using CPL parameterization, as is shown in figure \ref{fig-CPL}.
The upper panel of figure \ref{fig-CPL} shows reconstructed $w(z)$ and $f(z)$ from SnIa data alone.
From this panel, it is seen that $w$ and $f$ (i.e. DE density $\rho_{\Lambda}$) will rapidly decrease along with the increase of redshift $z$,
and the behavior of DE significantly deviate from the cosmological constant $\Lambda$ at 1$\sigma$ CL.
It should be mentioned that our results are consistent with some other works.
For example. Figure 2 of \cite{Shafieloo} shows that the best-fit value of $w_1$ is rather negative.
This means that the EOS $w$ of DE will also be quite negative at high-redshift region.

The lower panel of figure \ref{fig-CPL} shows reconstructed $w(z)$
and $f(z)$ from SnIa+BAO+CMB data. For this case, the result is
quite different, and the $\Lambda$CDM model is still consistent with
observational data at 1$\sigma$ CL. The difference from these two
panels of figure \ref{fig-CPL} shows the existence of the tension
among different types of observational data, which was recently
discussed in \cite{WeiHao}. Their result of the CPL parametrization
shows that for the Union compilation when CMB and BAO data is added
there is a $\Delta \chi^2=1.063$, while for the Constitution set
$\Delta \chi^2=4.846$, which is much larger. Their result is
consistent with ours. Besides, in \cite{ylwu} the authors get $(w_0,
w_1)=(-0.95^{+0.45}_{-0.18},0.41^{+0.79}_{-0.96})$ for the CPL
parametrization from SnIa+BAO data, which shows the existence of
tension between SnIa and BAO data compared with the upper pannel of
figure \ref{fig-CPL}.

As mention above, for fitting SNIa data alone,
we find that both $w$ and $\rho_{\Lambda}$ will rapidly decrease along with the increase of redshift $z$ by using CPL parameterization.
This result implies that the behavior of dark energy might be very different in different slices of redshifts,
and inspires us to separate redshifts into several bins and to investigate the piecewise constant $w$ model and the piecewise constant $\rho_{\Lambda}$ model.

 \

\subsection{$\Lambda$CDM2 and XCDM2 Model}

Let us discuss the cases in which redshifts are separated into 2bins.
Hereafter we will call 2 bins piecewise constant $w$ model as XCDM2 model,
and will call 2 bins piecewise constant $\rho_{\Lambda}$ model as $\Lambda$CDM2 model.

\begin{figure}
\centerline{\includegraphics[width=8cm]{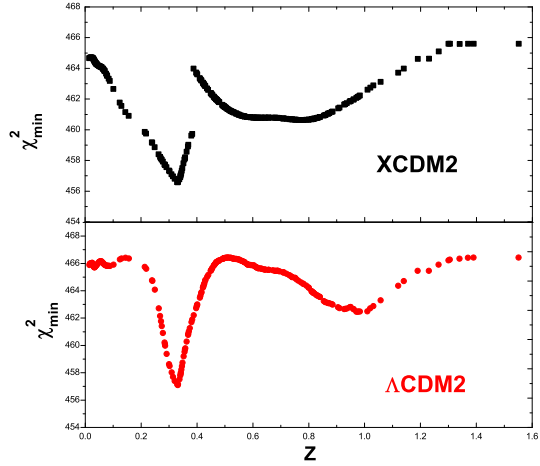}}
\caption{\label{fig1}
$\chi_{min}^{2}$ versus redshift $z$ for XCDM2 model and for $\Lambda$CDM2 model.
Only the Constitution SNIa sample is used in the analysis.}
\end{figure}

Figure \ref{fig1} shows the $\chi_{min}^{2}$ versus redshift $z$ for XCDM2 and $\Lambda$CDM2 models,
where only the Constitution SNIa sample is used in the analysis.
It is seen that the curves of $\chi_{min}^{2}$ of these two models are very similar,
and both these two models achieve their minimal $\chi_{min}^{2}$ at the discontinuity points of redshift $z_{1}=0.331$.
We will analyze the XCDM2 and $\Lambda$CDM2 model more explicitly at this discontinuity point.
The result is as follows.

For XCDM2 model, using the SNIa data and the discontinuity point
$z_{1}=0.331$, we find that the best-fit value and Corresponding
1$\sigma$ CL of the model parameters are
$\Omega_{m0}=0.466_{-0.032}^{+0.037}$,
$w_{0}=-1.118_{-0.418}^{+0.342}$, and $w_{1}$ with the best-fit
value $-508.170$ and the upper 1$\sigma$ CL value $-15.2$. For this
model, $\chi_{min}^{2}=456.574$ (As a comparison, for the CPL
parametrization to the same data, the $\chi_{min}^{2}=461.254$. The
XCDM2 model lower the $\chi_{min}^2$ by 4.32). Notice that the EOS
of the second redshift slice $w_{1}\ll -1$, this means the DE
density $\rho_{\Lambda}$ should decay very quickly at the range
$z>0.331$.

\begin{figure}
\centerline{\includegraphics[width=0.6\textwidth]{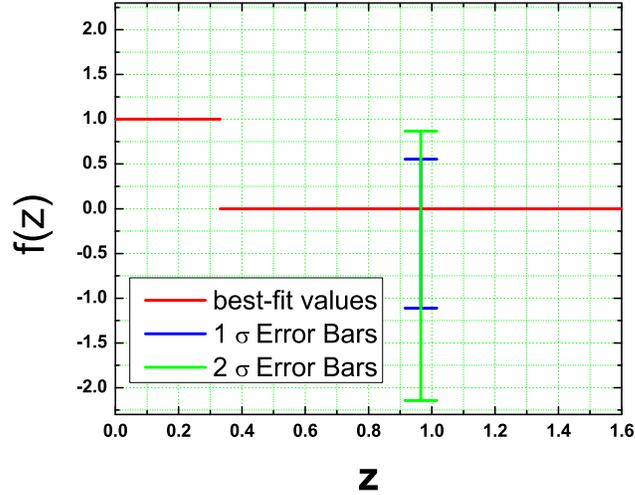}}
\caption{\label{fig2} Estimates of $\rho_{\Lambda}$/$\rho_{\Lambda0}$ of the $\Lambda$CDM2 model.
Only the Constitution SNIa sample is used in the analysis.
This figure shows a deviation from the $\Lambda$CDM over 1$\sigma$ CL.}
\end{figure}

For $\Lambda$CDM2 model, using the SNIa data and the discontinuity
point $z_{1}=0.331$, we find that the best-fit value and
Corresponding 1$\sigma$ CL of the model parameters are
$\Omega_{m0}=0.461_{-0.092}^{+0.128}$ and $\epsilon_{2}=(5.407\times
10^{-7})_{-0.462}^{+0.348}$, corresponding to
$\chi_{min}^{2}=456.893$. Notice that the best-fit $\Omega_{m0}$ and
the $\chi_{min}^{2}$ is very close to the result of the $XCDM2$
model. We also plot the 1$\sigma$ and 2$\sigma$ error bars for the
$\Lambda$CDM2 model in figure \ref{fig2}. It is seen that there is a
deviation from cosmological constant $\Lambda$ over 1$\sigma$
confidence level, which is similar to the results of CPL
parameterization and XCDM2 model.

\begin{table} \caption{Three toy models that mimic the step
function}
\begin{center}
\label{table1}
\begin{tabular}{cccc}
\hline\hline
~~~Model~~~ & ~~~Toy1~~~ & ~~~Toy2~~~ & ~~~Toy3~~~ \\
\hline\hline
~~~$f(z)$~~~ & ~~~{\Large$\frac{1+\xi^{-1}}{\xi^{-1}+\xi^{-3z}}$}~~~ & ~~~{\Large$\frac{1+\xi^{-1}}{\xi^{-1}+\xi^{-sz}}$}~~~ & ~~~{\Large$\frac{1}{1+e^{(z-z_{0})/v}}$}~~~ \\
\hline
~~~$\chi_{min}^{2}$~~~ & ~~~$457.805$~~~ & ~~~$457.499$~~~ & ~~~$456.771$~~~ \\
\hline\hline
\end{tabular}
\end{center}
\end{table}

\begin{figure}
\centerline{\includegraphics[width=8cm]{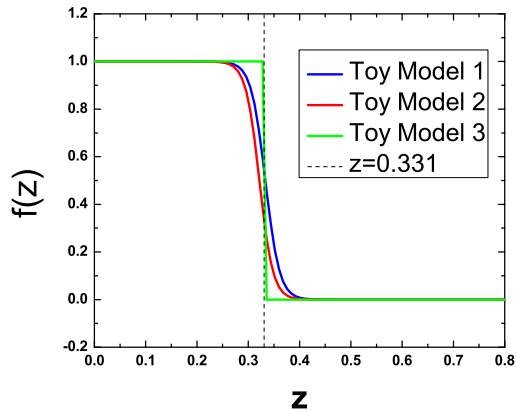}}
\caption{\label{fig3}
The evolution of DE density $\rho_{\Lambda}$ of three toy models.}
\end{figure}

To further verify this conclusion, we construct three toy models to
mimic the step function, as is listed in table \ref{table1}, and
analyze them using the SNIa data. It is found that,
for toy model 1, the best-fit model parameters are $\Omega_{m0}=0.457$ and $\xi=3.734\times 10^{-11}$,
corresponding to $\chi_{min}^{2}=457.805$;
for toy model 2, the best-fit model parameters are $\Omega_{m0}=0.462$, $\xi=2.644\times 10^{-11}$ and $s=3.108$,
corresponding to $\chi_{min}^{2}=457.499$;
for toy model 3, the best-fit model parameters are $\Omega_{m0}=0.462$,  $z_{0}=0.329$, and $v=4.514\times 10^{-11}$,
corresponding to $\chi_{min}^{2}=456.771$.
Notice that the numbers of free parameters of these three toy models are less than or as same as that of CPL parameterization,
so both these three models perform better than CPL parameterization in fitting the Constitution set.
we also plot the evolution of $f(z)$ of these three models in figure \ref{fig3},
where the best fit parameters of each model are adopted.
It is seen that all these three models have a sharp decrease in $\rho_{\Lambda}$ at $z\sim0.33$,
which is consistent with the $\Lambda$CDM2 model and the XCDM2 model.

\begin{figure}
\centerline{\includegraphics[width=8cm]{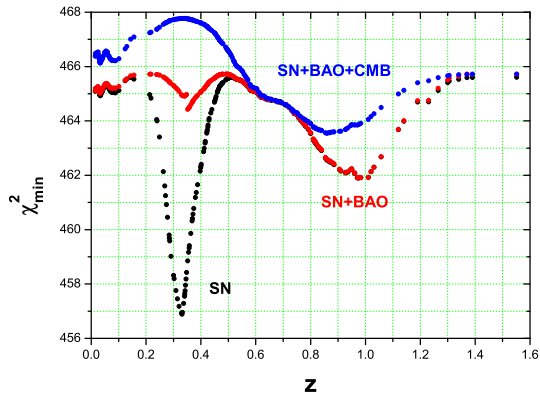}}
\caption{\label{fig4}
$\chi_{min}^{2}$ versus redshift $z$ for $\Lambda$CDM2 model,
where SNIa, SNIa+BAO, and SNIa+BAO+CMB data are used, respectively.}
\end{figure}

As is shown in figure \ref{fig4}, we also take the influence of BAO
and CMB data into account. Using the SNIa+BAO data, the best-fit
$\Lambda$CDM2 model has a discontinuity point $z_{1}=0.975$, and its
best-fit value and Corresponding 1$\sigma$ CL of the model
parameters are $\Omega_{m0}=0.277_{-0.025}^{+0.026}$ and
$\epsilon_{2}=7.813_{-6.298}^{+40.243}$, corresponding to
$\chi_{min}^{2}=461.906$. Using the SNIa+BAO+CMB data, the best-fit
parameters of $\Lambda$CDM2 model has a discontinuity point
$z_{1}=0.859$, and its best-fit value and Corresponding 1$\sigma$ CL
of the model parameters are $\Omega_{m0}=0.283_{-0.023}^{+0.024}$
and $\epsilon_{2}=1.677_{-0.732}^{+0.949}$, corresponding to
$\chi_{min}^{2}=463.552$. Therefore, although the step function
model in which $\rho_{\Lambda}$ rapidly decreases at redshift
$z\sim0.331$ is very powerful in fitting the SNIa data, it is not
the best model to fit the combination of SNIa, BAO and CMB data.
This shows the existence of the tension among different types of
observational data \cite{WeiHao}. This result is similar to the
result of CPL parameterization.

For a complete comparison, using SNIa, SNIa+BAO, and SNIa+BAO+CMB
data, we list $\chi_{min}^{2}$, $\chi_{min}^{2}/dof$, and $BIC$ for
XCDM2 model and $\Lambda$CDM2 model in table \ref{table2}, table
\ref{table3}, and table \ref{table4}, respectively. It is seen that
the $\chi_{min}^{2}$ of $\Lambda$CDM2 model are very close to those
of XCDM2 model. However, since the number of free parameters of
$\Lambda$CDM2 model is one fewer than that of XCDM2 model, the
$\chi_{min}^{2}/dof$ and $BIC$ of $\Lambda$CDM2 model are smaller
than those of XCDM2 model. This means the $\Lambda$CDM2 model
performs better than the XCDM2 model in fitting the observational
data. To make a further comparison, we also list the
$\chi_{min}^{2}$, $\chi_{min}^{2}/dof$, and $BIC$ for CPL
parameterization, respectively. One can see that, for SNIa,
SNIa+BAO, and SNIa+BAO+CMB data, the $\chi_{min}^{2}$ and $BIC$ of
$\Lambda$CDM2 model are 4.361, 3.534, and 2.549 smaller than those
of CPL parameterization, the $\chi_{min}^{2}/dof$ of $\Lambda$CDM2
model are 0.011, 0.009, and 0.006 smaller than those of CPL
parameterization. Therefore, the $\Lambda$CDM2 model presents a
significant improvement for the same data set obtained using the CPL
ansatz.

\begin{table}
\caption{The $\chi_{min}^{2}$ for XCDM2 model, $\Lambda$CDM2 model
and CPL parameterization}
\begin{center}
\label{table2}
\begin{tabular}{cccc}
  \hline\hline
  ~~~Model~~~ & ~~~SNIa~~~ & ~~~SNIa+BAO~~~ & ~~~SNIa+BAO+CMB~~~ \\
  \hline\hline
  ~~~XCDM2~~~ & ~~~$456.574$~~~ & ~~~$462.500$~~~ & ~~~$464.775$~~~ \\
  \hline
  ~~~$\Lambda$CDM2~~~ & ~~~$456.893$~~~ & ~~~$461.906$~~~ & ~~~$463.552$~~~ \\
  \hline
  ~~~CPL~~~ & ~~~$461.254$~~~ & ~~~$465.440$~~~ & ~~~$466.100$~~~ \\
  \hline\hline
\end{tabular}
\end{center}
\end{table}

\begin{table}
\caption{The $\chi_{min}^{2}/dof$ for XCDM2 model, $\Lambda$CDM2
model and CPL parameterization}
\begin{center}
\label{table3}
\begin{tabular}{cccc}
  \hline\hline
  ~~~Model~~~ & ~~~SNIa~~~ & ~~~SNIa+BAO~~~ & ~~~SNIa+BAO+CMB~~~ \\
  \hline\hline
  ~~~XCDM2~~~ & ~~~$1.162$~~~ & ~~~$1.174$~~~ & ~~~$1.177$~~~ \\
  \hline
  ~~~$\Lambda$CDM2~~~ & ~~~$1.160$~~~ & ~~~$1.169$~~~ & ~~~$1.171$~~~ \\
  \hline
  ~~~CPL~~~ & ~~~$1.171$~~~ & ~~~$1.178$~~~ & ~~~$1.177$~~~ \\
  \hline\hline
\end{tabular}
\end{center}
\end{table}

\begin{table}
\caption{The $BIC$ for XCDM2 model, $\Lambda$CDM2 model and CPL
parameterization}
\begin{center}
\label{table4}
\begin{tabular}{cccc}
  \hline\hline
  ~~~Model~~~ & ~~~SNIa~~~ & ~~~SNIa+BAO~~~ & ~~~SNIa+BAO+CMB~~~ \\
  \hline\hline
  ~~~XCDM2~~~ & ~~~$480.510$~~~ & ~~~$486.446$~~~ & ~~~$488.731$~~~ \\
  \hline
  ~~~$\Lambda$CDM2~~~ & ~~~$474.845$~~~ & ~~~$479.865$~~~ & ~~~$481.519$~~~ \\
  \hline
  ~~~CPL~~~ & ~~~$479.206$~~~ & ~~~$483.399$~~~ & ~~~$484.068$~~~ \\
  \hline\hline
\end{tabular}
\end{center}
\end{table}

\

\subsection{$\Lambda$CDM3 and XCDM3 Model}

Then, let us turn to the cases in which redshifts are separated into 3 bins.
In table \ref{table5}, we list $\chi_{min}^{2}$ and $BIC$ for
3 bins piecewise constant $w$ (XCDM3) model and 3 bins piecewise
constant $\rho_{\Lambda}$ ($\Lambda$CDM3) model. Comparing with the
cases of 2 redshift bins, the reduction of $\chi_{min}^{2}$ is very
small, but the increase of $BIC$ is quite obvious. Again, one can
see that the $\chi_{min}^{2}$ of $\Lambda$CDM3 model are very close
to those of XCDM3 model, but the $BIC$ of $\Lambda$CDM3 model are
much less than that of XCDM3 model. Therefore, for same number of
bins, a piecewise constant $\rho_{\Lambda}$ model always performs
better than a piecewise constant $w$ model. This shows the advantage
of using $\rho_{\Lambda}$, instead of $w$, to probe the variation of
DE.

\begin{figure}
\centerline{\includegraphics[width=0.6\textwidth]{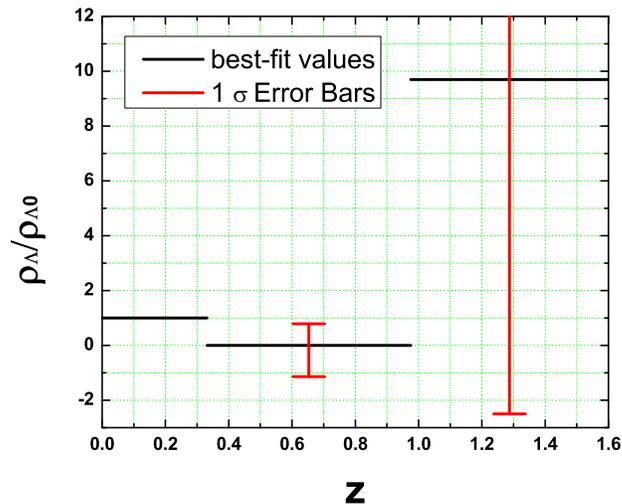}}
\caption{\label{LCDM3-rho} Estimates of
$\rho_{\Lambda}$/$\rho_{\Lambda 0}$ of the $\Lambda$CDM3 model. Only
the Constitution SNIa sample is used in the analysis. }
\end{figure}

We will not discuss the result of the $\Lambda$CDM3 and XCDM3 model
explicitly. Here we only mention that the second bins of the
$\Lambda$CDM3 and XCDM3 models also show a sharp decrease in
$\rho_{\Lambda}$, as is consistent with the CPL parameterization,
the LCDM2 model and the XCDM2 model. We get the minimal
$\chi^2_{min}=455.112$ at two discontinuity points $z=0.331$ and
$z=0.975$, and plot best-fit values and 1$\sigma$ CL of
$\rho_{\Lambda}$/$\rho_{\Lambda 0}$ for the $\Lambda$CDM2 model in
figure \ref{LCDM3-rho}.  As is consistent with figure \ref{fig-CPL}
, figure \ref{fig3} and figure \ref{fig2}, this figure also shows a
visible decrease of $\rho_{\Lambda}$ at $z>0.331$ and a deviation
from the $\Lambda$CDM in 1$\sigma$ CL. Notice that the Error Bar of
the third bin is too big to get any convincing information, the
reason is that there are only 20 supernovaes in this bin.

\begin{table} \caption{The $\chi_{min}^{2}$~($BIC$) for XCDM3
model and $\Lambda$CDM3 model}
\begin{center}
\label{table5}
\begin{tabular}{cccc}
  \hline\hline
  ~~~Model~~~ & ~~~SNIa~~~ & ~~~SNIa+BAO~~~ & ~~~SNIa+BAO+CMB~~~ \\
  \hline\hline
  ~~~XCDM3~~~ & ~~~$455.713~~(491.617)$~~~ & ~~~$459.598~~(495.517)$~~~ & ~~~$462.028~~(497.962)$~~~ \\
  \hline
  ~~~$\Lambda$CDM3~~~ & ~~~$455.112~~(485.032)$~~~ & ~~~$460.109~~(490.041)$~~~ & ~~~$461.059~~(491.004)$~~~ \\
  \hline\hline
\end{tabular}
\end{center}
\end{table}

\section{Summary}

In this work, we explore the cosmological consequences of the
recently released Constitution sample of 397 SNIa. By revisiting the
CPL parameterization, we find that, for fitting the Constitution set
alone, the behavior of DE significantly deviate from the
cosmological constant $\Lambda$, where $w$ and $\rho_{\Lambda}$ will
rapidly decrease along with the increase of redshift $z$. Inspired
by this clue, we separate the redshifts into different bins, and
discuss the models of a constant $w$ or a constant $\rho_{\Lambda}$
in each bin, respectively. It is found that for fitting the
Constitution set alone, $w$ and $\rho_{\Lambda}$ will also rapidly
decrease along with the increase of $z$, which is consistent with
the result of CPL model. Moreover, a step function model in which
$\rho_{\Lambda}$ rapidly decreases at redshift $z\sim0.331$ presents
a significant improvement ($\Delta \chi^{2}=-4.361$) over the CPL
parameterization, and performs better than other DE models. We also
construct three toy models that mimic this process, and find a DE
model that close to the step function model will be favored by the
Constitution SNIa data. By plotting the error bars of the piecewise
constant DE density of $\Lambda$CDM2 and $\Lambda$CDM3 model, we
show that there is a deviation from cosmological constant $\Lambda$
at about 1$\sigma$ confidence level; this may arise from some
biasing systematic errors in the handling of SNIa data, or more
interestingly from the nature of DE itself. In addition, for models
with same number of redshift bins, a piecewise constant
$\rho_{\Lambda}$ model always performs better than a piecewise
constant $w$ model; this shows the advantage of using
$\rho_{\Lambda}$, instead of $w$, to probe the variation of DE.

\section*{Acknowledgements}
We would like to thank Yun-Gui Gong, Bin Wang, Yi Wang, Hao Wei,
Yong-Shi Wu, Yue-Liang Wu, Xin Zhang and the anonymous referee for
kind help. This work is supported by the NSFC grant
No.10535060/A050207, a NSFC group grant No.10821504 and Ministry of
Science and Technology 973 program under grant No.2007CB815401.


\end{document}